\documentstyle[12pt,epsfig]{article}

\input epsf

\newcommand{\frat}[2]{\frac{\textstyle #1}{\textstyle #2}}

\newcommand{\vf}[1]{\mbox{\boldmath $#1$}}

\begin{document}
\begin{center}
{\Large {\bf 
Diquark condensate and quark interaction\\ with instanton liquid}}\\
\vspace{0.5cm}
S. V. Molodtsov, G. M. Zinovjev$^\dagger$\\
\vspace{0.5cm}
{\small\it State Research Center,
Institute of Theoretical and Experimental Physics,
117259, Moscow, RUSSIA}\\ 
$^\dagger$
{\small \it
Bogolyubov Institute for Theoretical Physics,\\
National Academy of Sciences of Ukraine, 
UA-03143, Kiev, UKRAINE}
\end{center}
\vspace{0.5cm}
\begin{abstract}
The interaction of light quarks and instanton liquid is analyzed at finite 
density of quark/baryon matter and in the phase of nonzero values of diquark
(colour) condensate. It is shown that instanton liquid perturbation produced
by such an interaction results in an essential increase of the critical value 
of quark chemical potential $\mu_c$ which provokes the perceptible increase of
quark matter density around the expected onset of the colour 
superconductivity phase.   
\end{abstract}
\vspace{0.5cm}

\\
\\

\vspace{0.5cm}
\noindent
The intriguing history of studying the structure of strongly interacting matter
abounded in many interesting but somewhat contradictory phenomenological 
results up to the advent of quantum chromodynamics (QCD). Since the time of its
discovery the possible existence of quark-gluon plasma (QGP) phase is one of 
the most fundamental predictions of QCD. This phase is phenomenologically
understood as a matter state where quarks and gluons being under extreme 
conditions of high temperature (T) and(or) quark/baryon density (driven by the
corresponding chemical potential $\mu$) are able to propagate relatively freely
over the considerable (of course, on the hadron scale) distances. Even the
simplest quantitative estimates of QGP physical characteristics were so
verysimilar that they allowed to launch the process of QGP experimental study 
in ultrarelativistic heavy ion collisions. And although the first series of
experiments at SPS CERN and AGS BNL have provided us with encouraging 
results a new generation of experiments is necessary to make convincing
conclusions. The physical programme for new experimental research certainly
needs serious quantitative analysis of the QCD phase diagram.

It seemed lattice QCD could be able to resolve the problem starting even on
the first theoretical principles. However, up to now this approach was 
effective and apparently reliable in producing the results mainly around the 
temperature axis of the ($\mu$---T)-plane and, unfortunately, relies on the
field theory formulation in imaginary time. Nevertheless, lattice QCD has
corraborated the phenomenological expectations that its phase structure is 
quite rich \cite{karsh}. In particular, nowadays we know that the chiral 
symmetry of initial QCD Lagrangian becomes broken and the quarks are getting
to be confined when the temperature decreases below a critical value. Moreover,
the lattice calculations show us that both physical phenomena result from the
corresponding phase transitions and their critical temperatures practically  
coincide. The latter observation indicates that both fundamental phenomena
could be tightly knitted and it should heuristically play an important role to
illuminate the confinement mechanism. 

The features of the QCD matter at high baryon densities remains poorly
understood in spite of the recent splash of renewed activity \cite{0}. One of
the major theoretical reasons comes from serious technical difficulties of
lattice simulations with Monte-Carlo techniques due to the complex fermion
determinant in QCD at finite quark densities. From the phenomenological side
the recent experimental data are not very informative for developing the 
theory in this direction. It is clear that present and future experiments with 
heavy ions at higher and higher energies should produce the strong interacting 
matter rather at higher temperature and closer to the temperature axis of the 
($\mu$---T)-plane. In such a situation the phenomenological treatment of 
studying along the $\mu$-axis should basically rely on the astrophysical 
observations of the compact stars. In the meantime, it was demonstrated that 
the diquark pairing in colour anti-triplet attractive channel induced by the 
instanton interaction vertex should be much more effective than an one-gluon 
exchange and at sufficiently large quark/baryon densities and small T could 
lead to colour superconductivity and a sophisticated picture of the QCD phase 
diagram \cite{0}. 

The present paper continues our preceding one \cite{mz} where the light quark
interactions with the instanton liquid (IL) at nonzero values of quark/baryon
chemical potentials were investigated in the phase of broken chiral symmetry
within the improved method of calculating the generating function of the IL
theory \cite{we0}. The new treatment of the functional integral is mainly
related to take into account the IL back-reaction upon the quark presence
which was always considered negligible. In fact, this effect is pretty weak
and should manifest itself in the leading order of expansion in the effective
coupling. Nevertheless, it was qualitatively argued that the interaction of IL
and quarks could increase the quark matter density around the onset of
expected colour superconductivity. As known \cite{diakcar}, \cite{rapp} this
density occurred to be surprisingly small ($n_q\sim 0.062~{\mbox{fm}}^{-3}$) 
and, hence, the diquark phase would come to play already at the density of
nuclear matter ($n_n\simeq 0.45~{\mbox{fm}}^{-3}$ and quark matter density is 
taken to be three times larger than the conventional nuclear one).
 
Here we obtain the quantitative estimate of the corresponding critical
chemical potential $\mu_c$ analyzing the system of the Gorkov equations for
the colour superconductor which we further calculate. We do not introduce any
simplification of the initial Lagrangian trying to deal with the 'exact'
four-quark interaction which is generated by (anti-)instantons. Such an
intention is simply compulsory because it is dictated by the precision of
problem where the effects of the IL perturbation are expected to be of the 
same order of magnitude compared to the terms which could usually be  
ignored.
 
Let us point out that in the IL approach \cite{2} the generating functional has
the following factorized form
$$
{\cal Z}~=~{\cal Z}_g~\cdot~{\cal Z}_\psi~.
$$
Here the first factor provides us with information on the gluon condensate
whereas the fermion factor ${\cal Z}_\psi$ serve to describe the quark
practice in the instanton environment \cite{diakcar}, \cite{2}. In what follows
we use the notations of Ref. \cite{mz} where the dimensionless variables 
(motivated by the form of interquark interaction kernel) were introduced, for
example, for the chemical potential it looks like $\mu\to \mu\bar\rho/2$ and
for the momenta $k_i\bar\rho/2 \to k_i,~i=1, \dots, 4$ where $\bar\rho$ is
the average size of pseudoparticle (PP). The quark determinant ${\cal Z}_\psi$ 
may be transformed with the auxiliary integration over the parameter $\lambda$ 
to the following form (for the colour $SU(3)$ group with the quarks of two  
flavours $N_f=2$):
\begin{eqnarray}
\label{1}
&&{\cal Z}_\psi\simeq\int d\lambda~\int D\psi^\dagger D\psi~\exp\left\{
n\bar\rho^4\left(\ln\frat{n\bar\rho^4}{\lambda~N}-1\right)\right\}
\times\nonumber\\
&&\times\exp\left\{\int \frat{dk}{\pi^4}~\sum_{f=1}^{2}\psi^\dagger_{f}(k)
~2~(-\hat k-i\hat\mu)\psi_{f}(k)+V\right\}~,
\\
&&V=2\lambda~(\psi^{\dagger L}_1~L_1~\psi_1^{L})
(\psi^{\dagger L}_2~L_2~\psi_2^{L})+
2\lambda~(\psi^{\dagger R}_1~R_1~\psi_1^{R})
(\psi^{\dagger R}_2~R_2~\psi_2^{R})~,\nonumber
\end{eqnarray}
where $\psi_f^{T}=(\psi_f^{R},\psi_f^{L}),~f=1,2$ are the quark fields with the
spinors of definite chirality, 
$\psi_f^{L,R}=P_\pm~\psi_f,~P_\pm=\frat{1\pm\gamma_5}{2}$, $n$ is the IL
density, $\mu_\nu=({\vf 0},\mu)$ and $N$ is the normalizing factor. For clarity
we take it to be equal a unity but, in principle, it could play a role of free
model parameter. This factor is inessential for the models with the fixed
value of the packing fraction parameter $n\bar\rho^4$ but in the model where 
it is admissible for variation the weak logarithmic dependence on N appears. 
The factors $2$ in Eq. (\ref{1}) come with making use of the dimensionless
variables. The term of four-fermion quark interaction $V$ may be directly
expressed with the chiral components as
$$(\psi^{\dagger L}_f~L_f~\psi_f^{L})=
\int \frat{dp_f dq_f}{\pi^8}~\psi^{\dagger L}_{f\alpha_f i_f}(p_f)~
L_{\alpha_f i_f}^{\beta_f j_f}(p_f,q_f;\mu)~\psi_f^{L\beta_f j_f}(q_f)~,$$
it being known that for the right hand fields the substitution $L \to R$
should be performed. The kernels $L_{\alpha_f i_f}^{\beta_f j_f}$ are defined
by the functions $h_i,~i=1, \dots, 4$ and the zero modes (the solutions of the 
Dirac equation with the chemical potential $\mu$ in the PP field) 
\begin{eqnarray}
h_4(k_4,k;\mu)&=&\frat{\pi}{4 k}
\{(k-\mu-ik_4)[(2k_4+i\mu)f^{-}_{1}+i(k-\mu-ik_4)f^{-}_2]+\nonumber\\
&+&(k+\mu+ik_4)[(2k_4+i\mu)f^{+}_{1}-i(k+\mu+ik_4)f^{+}_2]\}~,\nonumber
\end{eqnarray}
\begin{eqnarray}
h_i(k_4,k;\mu)&=&\frat{\pi~k_i}{4 k^2}
\left\{(2k-\mu)(k-\mu-ik_4)f^{-}_{1}+(2k+\mu)(k+\mu+ik_4)f^{+}_1+
\right.\nonumber\\
&+&\left[2(k-\mu)(k-\mu-ik_4)-\frat{1}{k}(\mu+ik_4)[k_4^{2}+(p-\mu)^2]
\right]f^{-}_{2}+\nonumber\\
&+&\left.
\left[2(k+\mu)(k+\mu+ik_4)+\frat{1}{k}(\mu+ik_4)[k_4^{2}+(p+\mu)^2]
\right]f^{+}_{2}\right\}~,\nonumber
\end{eqnarray}
where $k=|{\vf k}|$ if the spatial components of 4-vector $k_\nu$ are
considered and
$$f_1^{\pm}=
\frat{I_1(z^\pm)K_0(z^\pm)-I_0(z^\pm)K_1(z^\pm)}
{z^\pm}~,\\
f_2^{\pm}=\frat{I_1(z^\pm)K_1(z^\pm)}{z^2_{\pm}}~,~~
z^\pm=\frat{\rho}{2}\sqrt{k_4^{2}+(k\pm\mu)^2}~,
$$
with the modified Bessel functions $I_i,~K_i~(i=0,1)$.
Let us introduce the scalar function $h(k_4,k;\mu)$ related to 
three-dimensional component 
$h_i(k_4,k;\mu)=h(k_4,k;\mu)~\frat{k_i}{k}~,i=1,2,3$
(we omit the arguments of functions $h_i$ when it does not mislead).
$$L_{\alpha i}^{\beta j}(p,q;\mu)=S^{i k}(p;\mu)
\epsilon^{k l}~U^{\alpha}_{l}~U^{\dagger\sigma}_{\beta}
\epsilon^{\sigma n}S^{+}_{n j}(q;-\mu)~,
$$
$S(p;\mu)=(p+i\mu)^{-}~h^{+}(p;\mu)~,~~
S^+(p;-\mu)=\stackrel{*}h\!\!{^{-}}(p;-\mu)(p+i\mu)^{+}$ where it is valid 
for the conjugated function $\stackrel{*}h_\mu(p;-\mu)=h_\mu(p;\mu)$ and
$\epsilon$ is an antisymmetric matrix with $\epsilon_{12}=-\epsilon_{21}=1$. 
Here $p^\pm$ and other similar designations are used for the four-vectors
spanned by $\sigma^\pm_{\nu}$-matrices, 
$\sigma^{\pm}_\mu=(\pm i{\vf \sigma},1),$ (${\vf \sigma}$ is the three-vector
of the Pauli matrices), for example, $p^\pm=p^\nu\sigma^\pm_{\nu}$ and
$U$ is a matrix of rotations in the colour space.
The similar relations are valid for the right hand components 
$$(\psi^{\dagger R}_f~R_f~\psi_f^{R})=
\int \frat{dp_f dq_f}{\pi^8}~\psi^{\dagger R}_{f\alpha_f i_f}(p_f)~
R_{\alpha_f i_f}^{\beta_f j_f}(p_f,q_f;\mu)~\psi_f^{R\beta_f j_f}(q_f)~,$$
with the kernel
$$R_{\alpha i}^{\beta j}(p,q;\mu)=T^{ik}(p;\mu)
\epsilon^{k l}~U^{\alpha}_{l}~U^{\dagger\sigma}_{\beta}
\epsilon^{\sigma n}T^{+}_{n j}(q;-\mu)~,
$$ 
where $T(p;\mu)=(p+i\mu)^{+}~h^{-}(p;\mu)~,~~
T^+(p;-\mu)=\stackrel{*}h\!\!{^{+}}(p;-\mu)(p+i\mu)^{-}$.
The components of matrices $(p+i\mu)^{\pm}$ and $h^{\mp}(p;\mu)$ commute
because the vector-function ${\vf h}(p)$ is spanned by the vector ${\vf p}$
only. Then the following identities are easily understood
$$T(p;\mu)=S^{+}(p;-\mu)~,~~T^+(p;-\mu)=S(p;\mu)~.$$
In what follows we omit the $\mu$-dependence of matrices $T,~T^+$ as the 
chemical
potential enters the matrix $T$ being always positive and the matrix $T^+$ 
being negative only. Besides, two other identities would also be helpful  
$$\sigma_2 T^{T}(p)\sigma_2=T^+(p)~,~~
\sigma_2 T^{+ T}(p)\sigma_2=T(p)~,
$$
where $T^T$ means a transposed matrix. 

Dealing with the following averages interesting to study the diquark
condensates (\ref{1}) 
$$
\langle\psi^{L,R}_{1\alpha i}(p)\psi^{L,R}_{2\beta j}(q)\rangle=
\epsilon_{12}~\epsilon_{\alpha\beta}~\pi^4~\delta(p+q)~F^{L,R}_{ij}(p)~,
$$
$$
\langle\psi^{L}_{f\alpha i}(p)\psi^{\dagger R}_{g\beta j}(q)\rangle=
\delta_{fg}~\delta_{\alpha\beta}~\pi^4~\delta(p-q)~G^{LR}_{ij}(p)~,
$$
and making use of the effective action of Eq. (\ref{1}) one may obtain the 
Gorkov equations similar to ones treated in Ref. \cite{rapp}
\begin{eqnarray}
\label{2}
&&\left[G_{0}^{+}(p)\right]^{-1}~F^L(p)-\Sigma^R(p)~G^{LR~T}(-p)=0~,
\nonumber\\ 
&&\left[G_{0}^{+}(p)\right]^{-1}~G^{LR}(p)-
\Sigma^R(p)~F^{+R~T}(p)=1~,
\nonumber\\
[-.2cm]
\\[-.25cm]
&&\left[G_{0}^{-}(p)\right]^{-1}~F^R(p)-
\Sigma^L(p)~G^{RL~T}(-p)=0~,\nonumber\\
&&\left[G_{0}^{-}(p)\right]^{-1}~G^{RL}(p)-
\Sigma^L(p)~F^{+L~T}(p)=1~,\nonumber
\end{eqnarray}
where $\left[G_0^{\pm}(p)\right]^{-1}=-2~(p+i\mu)^\pm$ means the free Green
function, $\Sigma^{R}(p)=\Delta^{R}~T(p)~\epsilon~T^{T}(-p)$,
$\Sigma^{L}(p)=\Delta^{L}~T^{+}(p)~\epsilon~T^{+T}(-p)$ and
$\Delta^{L,R}$ denotes a gap. The form of $\Sigma$-matrices results from the
kernel structure of equations if averaging over colour orientations done
(remember that we consider colour stochastic ensemble). In order to complete
Eqs. (\ref{2}) we need the following gap equations 
$$\epsilon \Delta^{R}=\frat{2\lambda}{N_c(N_c-1)}~
\int \frat{dq}{\pi^4} \left[T^{+}(q)~F^{R}(q)~T^{+T}(-q)-
T^{+}(-q)~F^{R~T}(q)~T^{+T}(q) \right]~,
$$
$$\epsilon \Delta^{L}=\frat{2\lambda}{N_c(N_c-1)}~
\int \frat{dq}{\pi^4} \left[T(q)~F^{L}(q)~T^{T}(-q)-
T(-q)~F^{L~T}(q)~T^{T}(q) \right]~.
$$
Apparently, the right hand sides of these equations are proportional $\epsilon$
because they are the differences a matrix and its transposed form. Similar
equations are valid for the conjugated matrices
\begin{eqnarray}
\label{3}
F^{+L~T}(p)&\left[G_{0}^{-}(p)\right]^{-1}&-G^{RL~T}(-p)~\Sigma^{+R}(p)=0~,
\nonumber\\
G^{RL}(p)&\left[G_{0}^{-}(p)\right]^{-1}&-
F^R(p)~\Sigma^{+R}(p)=1~,
\nonumber\\
[-.2cm]
\\[-.25cm]
F^{+R~T}(p)&\left[G_{0}^{+}(p)\right]^{-1}&-
G^{LR~T}(-p)~\Sigma^{+L}(p)=0~,\nonumber\\
G^{LR}(p)&\left[G_{0}^{+}(p)\right]^{-1}&-
F^L(p)~\Sigma^{+L}(p)=1~,\nonumber
\end{eqnarray}   
with the corresponding gap equations 
$$\epsilon \Delta^{+R}=\frat{2\lambda}{N_c(N_c-1)}~
\int \frat{dq}{\pi^4} \left[T^{T}(-q)~F^{+R~T}(q)~T(q)-
T^{T}(q)~F^{+R}(q)~T(-q) \right]~,
$$
$$\epsilon \Delta^{+L}=\frat{2\lambda}{N_c(N_c-1)}~
\int \frat{dq}{\pi^4} \left[T^{+T}(-q)~F^{+L~T}(q)~T^{+}(q)-
T^{+T}(q)~F^{+L}(q)~T^{+}(-q) \right]~,
$$
where $\Sigma^{+R}(p)=\Delta^{+R}~T^{+T}(-p)~\epsilon~T^{+}(p)$ and
$\Sigma^{+L}(p)=\Delta^{+L}~T^{T}(-p)~\epsilon~T(p)$.
In this paper we limit ourselves to treating the diquark condensate only. 
However,
as known \cite{rapp} the mixed phase of non-zero values both chiral and colour
condensates could exist at $\mu_c\sim 300~{\mbox{MeV}}$ realizing the 
transitional regime for the onset of colour superconducting phase. In order to
bring this phase to the play the equation system should be extended including
another averages as
$$
\langle\psi^{L,R}_{f\alpha i}(p)\psi^{\dagger L,R}_{g\beta j}(q)\rangle=
\delta_{fg}~\delta_{\alpha\beta}~\pi^4~\delta(p-q)~G^{LL,RR}_{ij}(p)~.
$$

From the Eqs. (\ref{2}) and (\ref{3}) we find
\begin{eqnarray}
G^{LR}(p)&=&G_0^{+}(p)+G_0^{+}(p)~\Sigma^R(p)~F^{+R~T}(p)~,
\nonumber\\
F^{+R~T}(p)&=&G^{LR~T}(-p)~\Sigma^{+L}(p)~G_0^{+}(p)~.
\nonumber
\end{eqnarray}
With the auxiliary matrices $C^{+L}(p)=\Delta^{+L}~T^{+}(-p)~T(p)$,
$C^{R}(p)=\Delta^{R}~T(p)~T^{+}(-p)$ we may rewrite the matrices $\Sigma$ as 
$$\Sigma^{+L}(p)=\epsilon~C^{+L}(p)~,~~\Sigma^{R}(p)=C^{R}(p)~\epsilon~.
$$
The identities for $T$-matrix mentioned above help to show the validity of
the relations
$$\epsilon~C^{+L~T}(-p)~\epsilon^T=C^{+L}(p)~,$$
$$\epsilon~C^{R~T}(-p)~\epsilon^T=C^{R}(p)~.$$
And for the Green function we have
$$G^{LR}(p)=G_0^{+}(p)+G_0^{+}(p)~C^{R}(p)~\epsilon~G^{LR~T}(-p)~\epsilon~
C^{+L}(p)~G_0^{+}(p)~.
$$
Combining the properties of $C$ matrices together with the identities for free
Green functions
$$\sigma_2~G_0^{\pm T}(p)~\sigma_2=~G_0^{\mp}(p)~,
$$
we are able to get the complete equation for calculating the function $G^{LR}$
in the form
$$\epsilon~G^{LR}(-p)~\epsilon^T=
G_0^{-}(-p)+G_0^{-}(-p)~C^{+L}(p)~G^{LR}(p)~C^{R}(p)~G_0^{-}(-p)~.$$
The vector-function ${\vf{ h}}(p)$ being spanned by the vector ${\vf{ p}}$
helps to conclude that all the matrices $G_0^{\pm}$, $C^{+L}$, $C^R$ commute
with each other. Then searching the solution for the Green function 
$G^{LR}$ by iterating one finds immediately that the Green function commutes 
with those matrices and finally obtains the equation to calculate it in the
form
$$G^{LR}(p)=[1+H(p)]~G_0^{+}(p)+H^2(p)~G^{LR}(p)~,$$
or
$$[1-H(p)]~G^{LR}(p)=G_0^{+}(p)~,
$$
where $H(p)=G_0^{+}(p)~C^R(p)~G_0^{-}(-p)~C^{+L}(p)$.

The structure of matrices $H(p)$ allows us to conclude that their sum
$$H(p)+H(-p)=\alpha(p)~$$
and product
$$H(p)~H(-p)=\beta(p)~,$$
are proportional to the unity matrices (it is clear by definition that
$\alpha(-p)=\alpha(p),~\beta(-p)=\beta(p)$).
 
Introducing notation $g_\nu=h_\nu(-p)$ we are able to present the functions 
$\alpha(p)$ and $\beta(p)$ in the compact form as
$$\alpha(p)=4~\Delta^R~\Delta^{+L}~[-4~A(p)~(hg)+2~(p^2+\mu^2)~(h^2)~(g^2)]~,
$$
where
$$A(p)=(p^2+\mu^2)~(hg)-2i~\mu~p(g_4 h-h_4 g)~.
$$
and a scalar product is naturally defined $(hg)=\sum_{i=1}^4 h_i g_i$
together with the functions $(h^2)$ and $(g^2)$ squared. Getting the last term
of $A(p)$ we used the designation of scalar function ${\vf h}$ mentioned above
and for the function $\beta$ we have
$$\beta(p)=16~\left(\Delta^R\right)^2~
\left(\Delta^{+L}\right)^2~\left(p^2+\mu^2\right)^2~(h^2)^2~(g^2)^2~.
$$

As both functions $\alpha$ and $\beta$ are spanned by the unit matrices the
solution for the Green function may be given in
$$G^{LR}(p)=\frat{G_0^{+}(p)~[1-H(-p)]}{1-\alpha(p)+\beta(p)}~.
$$
The gap equation then looks like
\begin{equation}
\label{gap}
\Delta^L=\frat{2\lambda}{N_c (N_c-1)}~\int\frat{dp}{\pi^4}~
\frat{\alpha(p)-2~\beta(p)}{\Delta^{+L}(1-\alpha(p)+\beta(p)}~.
\end{equation}
We are interested in the solution of the form
$\Delta^{R}=\Delta^{+L}=\Delta^{L}=\Delta^{+R}$ at $\lambda<0$,
which is dictated by the symmetries of four-quark interaction kernels. Let us
remember that any kernel for every quark sort traditionally carries the
imaginary unit factor $i$ \cite{2}, \cite{dp2}. The sign choice of $\lambda$ 
relies on this fact but, in principle, there is an alternative 
$\Delta^{R}=\Delta^{+L}=-\Delta^{L}=-\Delta^{+R}$ for $\lambda>0$, if the
kernels are redefined. An analysis shows that the denominator of 
Eq. (\ref{gap}) 
is always positive and, therefore, the solution of this equation does exist at
pretty large $\lambda$. 

The quark matter state at finite chemical potential is defined by the saddle 
point of functional Eq. (\ref{1}) which we treat further maintaining the
first nonvanishing contribution which is the figure-eight type diagram 
(see, for example, \cite{diakcar}), 
$$I=2~(N_c-1)~\int\frat{dp}{\pi^4}~
\frat{\alpha(p)-2~\beta(p)}{1-\alpha(p)+\beta(p)}~.
$$
In our consideration this contribution to the generating functional
($Z_\psi\sim e^{W}$) would occur to be
\begin{equation}
\label{GZ}
W=-n\bar\rho^4~\ln\lambda+\lambda~\langle Y\rangle~,
~~\langle Y\rangle\simeq I~.
\end{equation}
For the simplest situation of constant IL density the saddle point equation 
reads
$$n\bar\rho^4-\lambda~\langle Y\rangle=0~.
$$
Comparing it with the gap equation Eq. (\ref{gap}) one could notice the
peculiar feature which is very practical to keep numerical calculations under
control. The saddle point equation leads to the condition of gap independent
of $\mu$. 

It was demonstrated in Refs. \cite{we0}, \cite{we} that the quark backreaction
upon IL could be perturbatively estimated by studying the small variations of
the IL parameters $\delta n$ and $\delta \rho$ around their equilibrium values
$n$ and $\bar\rho$. Such variations are incorporated by the IL theory if the
deformable (crumpled) (anti~-~)\-in\-stan\-tons of size $\rho$, being 
the function of 
$x$ and $z$, i.e. $\rho\to\rho(x,z)$, are treated as the saturating
configurations of the functional integral. In addition, the variations of zero 
modes in the interaction vertices of quark determinant at the transformation
$\bar\rho \to \bar\rho+\delta\rho$ should be taken into account. Then as the
output we have that for the long wave length excitations (for example, 
$\pi$-mesons) the deformation field describes colourless scalar excitations of
IL with the mass gap $M$ of the order of several hundreds MeV,
$M^2=\frat{\nu}{\kappa}$ where $\nu=\frat{b-4}{2}$, $b=\frat{11~N_c-2~N_f}{3}$,
$\kappa$ is the kinetic coefficient being derived within the quasi-classical
approach. $N_c$ and $N_f$ are the numbers of quark colours and flavours,
respectively. Our estimates give for this coefficient value of a few single
instanton actions $\beta=8\pi^2/g^2$, i.e. $\kappa\sim c~\beta$ 
(with the factor $c\sim 1.5$ --- $6$ according to the ansatz taken for the 
saturating configurations) \cite{we}.

Then, besides the diagrams with four legs (see, the term $V$ in (\ref{1}))
the extra diagrams with the scalar field attached (relatively speaking, the
derivative in $\rho$ of the vertex function in which the variation of the
functions describing the zero mode  
\begin{equation}
\label{4}
h_i\to h_i+\frat{\partial h_i}{\partial \rho}~\delta\rho~,~~
i=1, \dots, 4~,
\end{equation}
should be performed) are generated. Due to the diquark condensate presence
handling the Lagrangian is substantially simplified if one
restricts oneself with the leading contributions coming from the tadpole
diagrams. Indeed, the leading contributions come from the term $V$ and the
term shown in Fig. 1 where two vertices are linked with the propagator
$\frat{1}{M^2}$ of scalar field. 

\begin{figure}[htb]
\centerline{\epsfig{file=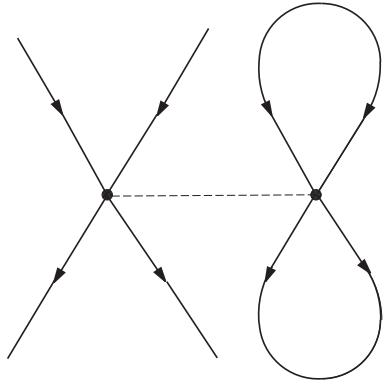,width=5.cm}}
\vspace*{-.1 cm}
\caption{The diagram of tadpole approach (see, the text). 
The solid (dashed) lines correspond to the fermion (scalar) field.}
\end{figure}

Analyzing the modified Lagrangian it is easy to see that the form of
the Gorkov equations would be the same if everywhere $\Sigma$ is meant as an
improved one $\Sigma+\delta\Sigma$ in which the correction $\delta\Sigma$ 
is constructed with the modified functions Eq. (\ref{4}). Moreover, the
result for the Green function is formally the same if the modified  
functions $\Sigma$ are meant. In particular, the form of gap equation is also
retained because its kernel includes again the differences of the matrices
which are constructed from the derivatives in $\rho$ of matrices $T,~T^{+}$ 
and their transposed forms. This difference is spanned by the unit vector as 
well. However, the following substitutions
$$\alpha(p)\to\alpha(p)+\delta\alpha(p)~,$$
$$\beta(p)\to\beta(p)+\delta\beta(p)~,$$
where
$$\delta\alpha(p)=J~\frat{\partial \alpha}{\partial\rho}~,$$
$$\delta\beta(p)=J~\frat{\partial \beta}{\partial\rho}~,$$
are supposed to be performed and here $J$ is the contribution of figure-eight 
type diagram with the scalar field propagator as an external leg 
$$J=\frat{2\lambda}{N_c(N_c-1)}~\frat{1}{n\bar\rho^4\kappa}
\frat{1}{4 M^2}~I~.
$$
Let us mention that effectively a dependence on the kinetic term $\kappa$ 
disappears (remember that $M^2=\frat{\nu}{\kappa}$) and its precise value is 
not operative in the approximation
developed. The derivative of the figure-eight type diagram which will shortly
be necessary to proceed looks like
$$\frat{\partial I}{\partial \rho}=2~(N_c-1)~\int\frat{dp}{\pi^4}~
\left\{\frat{1-\beta(p)}{(1-\alpha(p)+\beta(p))^2}~\delta \alpha(p)+
\frat{\alpha(p)-2}{(1-\alpha(p)+\beta(p))^2}~\delta \beta(p)\right\}~.
$$
And finally the modification of generating functional should be accommodated
bringing Eq. (\ref{GZ}) to the form 
$$W=-n\bar\rho^4~\left(\ln\frat{n\bar\rho^4}{\lambda}-1\right)+
\lambda~\langle Y\rangle~.$$
Then the new (if the variation of the IL density is absorbed) equation to 
calculate the saddle point reads
$$n\bar\rho^4-\lambda~(n\bar\rho^4)'\ln\frat{n\bar\rho^4}{|\lambda|}-
\lambda~\langle Y\rangle=0~.
$$
and the IL density is \cite{we0}
\begin{equation}
\label{ilden}
n\bar\rho^4=\frat{\nu}{2\beta\xi^2}+
\left[\left(\frat{\nu}{2\beta\xi^2}\right)^2+
\frat{\left(\frat{\delta I}{\delta\rho}\right)^{'}}{\beta \xi^2}~
\frat{\Gamma(\nu+1/2)}{2\sqrt{\nu}~\Gamma(\nu)}
\right]^{1/2}~,
\end{equation}
where the prime available means the derivative in $\lambda$, the constant 
$\xi^2=\frat{27}{4}\frat{N_c}{N_c^{2}-1} \pi^2$ 
is a measure of interaction in the stochastic ensemble of PPs with an average
size as\\ 
$\bar\rho\Lambda=\exp\left\{-\frat{2N_c}{2\nu-1}\right\}$. 
The derivative in $\lambda$ of the figure-eight type diagram looks like 
$$\left(\frat{\delta I}{\delta \rho}\right)^{'}=
4~(N_c-1)\frat{\Delta^{'}}{\Delta}~
\int\frat{dp}{\pi^4}~\left\{
\frat{1+\alpha-6\beta+\alpha\beta+\beta^2}
{(1-\alpha+\beta)^3}~\delta \alpha+
\frat{-4+7\alpha+4\beta-\alpha\beta-\alpha^2}
{(1-\alpha+\beta)^3}~\delta \beta\right\}~
$$
and the derivative $\Delta^{'}$ is defined by the following equation
$$\frat{N_c (N_c-1)}{2\lambda^2}=\frat{2\Delta^{'}}{\Delta^3}
\int\frat{dp}{\pi^4}~
\frat{2\beta-2\beta^2+2\alpha\beta-\alpha^2}
{(1-\alpha+\beta)^2}~.
$$

\begin{figure}[htb]
\centerline{\epsfig{file=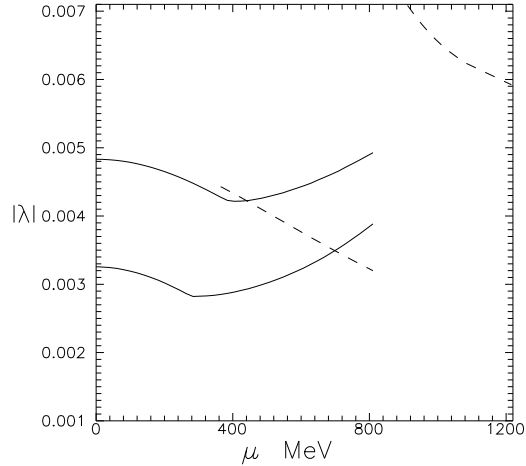,width=6.8cm}}
\vspace*{-.1 cm}
\caption{The saddle point $\lambda$ as the function of 
chemical potential $\mu$ at $N_c=3$ and $N_f=2$. 
The solid lines correspond to the phase of broken chiral symmetry (the lower
one absorbs the IL perturbation). The dashed lines correspond to the colour
superconductivity phase (the left hand one does not include the tadpole
contribution whereas the right hand one does.}
\end{figure}

Fig. 2 shows the results of calculating the parameter $\lambda$ (which is
proportional to the free energy within the precision of one loop approximation)
as the function of chemical potential at $N_c = 3, N_f = 2.$ The dashed lines
expose a behaviour in the phase of nonzero values of diquark condensate (the
left hand line (lower line) corresponds to the calculations if the quark 
interaction with 
IL is ignored) whereas the solid lines show the behaviours in the phase of 
broken chiral symmetry (upper line for the quark interaction with IL ignored)
 \cite{mz}. The saddle point parameter  
$\lambda_1$ of Ref. \cite{mz} and that exploited in the present paper are
related as 
$\lambda_1^{2}=-\frat{\lambda~n\bar\rho^4}{2(N_c-1)N_c}$.
The crossing points of the curves are fixing the onset of colour 
superconducting phase. As mentioned above, actually, there is a transitional 
region of mixed phase where both chiral and diquark condensates develop
non-zero magnitudes and the descent is not so steep. However, these details 
are inessential for further discussion. What is more interesting to be noticed 
comes from the crossing point of lower solid and dashed lines positioned
on the plot. 

\begin{figure}[htb]
\centerline{\epsfig{file=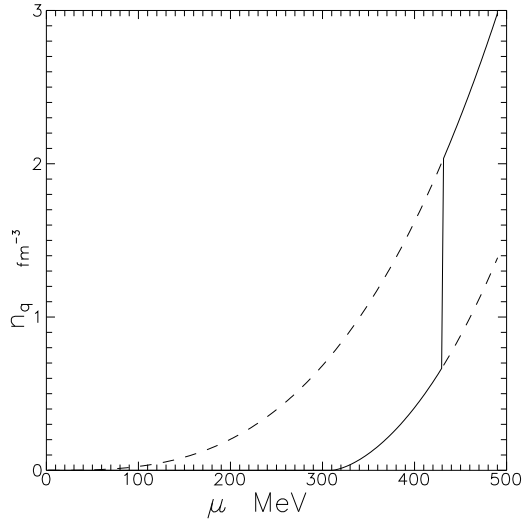,width=6.8cm}}
\vspace*{-.1 cm}
\caption{The quark density being defined by the crossing point of the upper
solid and left hand dashed lines in Fig. 2. The solid line corresponds to the 
stable phase and the dashed one corresponds to the metastable one.}
\end{figure}

The first crossing point (along the $\mu$-axis) which corresponds
to the approach without the tadpole contributions included looks to be slightly
larger in magnitude than that obtained earlier $\mu_c\simeq 300~{\mbox{MeV}}$ 
in Refs. \cite{diakcar}, \cite{rapp}. However, it does not signal about 
shifting the critical point to larger values of $\mu$ in the approach dealing
with the initial Lagrangian. The IL parameters of various approaches are
sometimes slightly different and, in principle, could be optimized fitting,
for example, $\Lambda_{QCD}$. Thus, the shift of $\mu_c$ value pointed out is
well within a precision of the IL theory. Regarding the second crossing point
(when IL is perturbed by quarks) the conclusion could be more indicative. Here
$\mu_á$ becomes significantly larger. The corresponding quark matter densities
as the functions of chemical potential for both approaches are depicted in
Fig. 3 and Fig. 4. Apparently, on both plots the density values corresponding
to the onset of colour superconductivity phase are noticeably larger than the
density of normal nuclear matter. Perhaps, our result could be taken as a
general indication on the considerable role of supplementary (insignificant
on the instanton background) interactions of light quarks. They are able to
change the estimate of critical density for colour diquark condensation
drastically.

\begin{figure}[htb]
\centerline{\epsfig{file=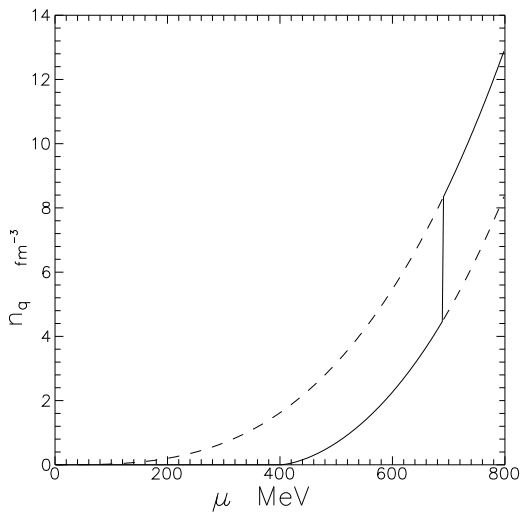,width=6.8cm}}
\vspace*{-.1 cm}
\caption{The quark density being defined by the crossing point of the lower
solid and left hand dashed lines in Fig. 2. The solid line corresponds to the 
stable phase and the dashed one corresponds to the metastable one.}
\end{figure}

Now we would like to comment on the crossing point of upper dashed and lower
solid lines which is not shown in Fig. 2. It corresponds the unrealistic values
of $\mu_á$ and in that region the IL is hardly applicable. One of the limits 
for IL approach coming from the very large values of chemical potential points
out that with quark matter density increasing the average interquark distances
may become very small and the 'Coulomb' (perturbative) field strengths would
occur to be comparable with the (anti-)instanton ones. Then the 
(anti-)instanton superposition is not a proper configuration to saturate the
functional integral. Thus, our conclusion resulting from the approach
developed gives a message that the quark perturbation of IL leads 
the corresponding curves to move apart inherent in the phases of non-zero 
values of diquark
and chiral condensates in Fig. 2. The 'chiral' curve becomes steeper but
'diquark' curve is displaced to the larger values of $\mu$ increasing its
critical value. It looks like giving more reliability to our understanding of
the perturbative fields role in the IL theory could provide us with more 
accurate estimates for the phenomena considered.

\begin{figure}[htb]
\centerline{\epsfig{file=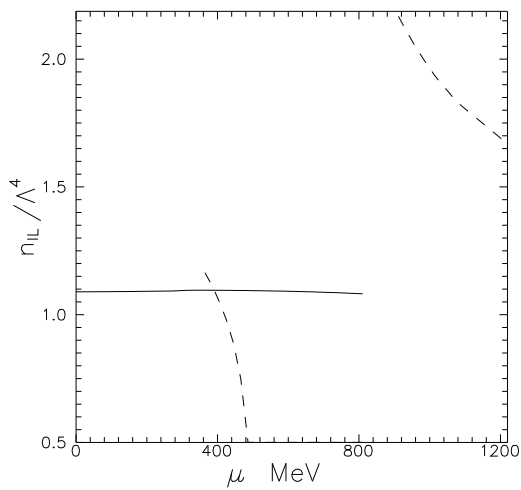,width=6.8cm}}
\vspace*{-.1 cm}
\caption{ The IL density for the approach at $N_c=3$ and $N_f=2$. The solid
line corresponds to the calculation in the phase of broken chiral symmetry.
Both dashed lines correspond to the calculation in the phase of non-zero
diquark condensate values. The lower dashed line gives the estimate of IL
density without the tadpole contributions included.
Regarding the upper dashed line see the text.}
\end{figure}

Fig. 5 demonstrates the IL density as the function of $\mu$ when the quark
backreaction is incorporated in the phase of broken chiral symmetry (solid
curve) and in the phase of non-zero values of diquark condensate (dashed
curve). It is interesting to notice that the curves manifest the different
character of quark influence on IL. In the phase of broken chiral symmetry
the density is almost constant in the suitable interval of $\mu$ whereas in
the colour superconducting phase quickly disappears. It is interesting that
for the former the corresponding analogue of the tadpole contribution
$\left(\frat{\delta I}{\delta\rho}\right)^{'}$ in Eq. (\ref{ilden}) is strictly
positive and, therefore, can lead to the increase of the IL density (gluon
condensate) which is truely insignificant and demonstrates simply the approach
sensitivity to the dynamical quark mass variation  \cite{mz}. In the latter
case the sign of tadpole contribution is changing. 
As seen from Fig. 5 it leads
to the drastic change of the IL density behaviour and the gluon condensate is
getting weaker. Strictly speaking this effect has already been discussed in 
\cite{kerb}.

\noindent
The authors are partly supported by STCU\#015c,
CERN-INTAS 2000-349, NATO~2000-PST.CLG 977482 Grants. S.V.M. 
expresses his gratitude to Dr. P. Giubellino and 
Dr. P. Kuijer for interesting discussions of the QGP physics and support.
Authors also indebted to the Faberge Fund for providing us with the excellent 
working conditions.

\end{document}